\documentclass[useAMS,usenatbib]{mn2e}
\usepackage{latexsym,hhline,epsfig,longtable}
\usepackage{graphicx,color}
\usepackage{ulem}

\def\chem#1{$^{#1}$}  

\title[Final fates of super- and massive AGB stars]{Super and massive AGB stars - IV. Final fates - Initial to final mass relation}
\author[C.L. Doherty, P. Gil-Pons, L. Siess, J.C. Lattanzio, H.H.B. Lau]{Carolyn L. Doherty$^{1}$\thanks{E-mail:carolyn.doherty@monash.edu}, Pilar Gil-Pons$^{2}$, Lionel Siess$^{3}$, John C. Lattanzio$^{1}$, \newauthor and Herbert H.B Lau$^{4}$\\ 
$^{1}$Monash Centre for Astrophysics (MoCA), School of 
Mathematical Sciences, Monash University, Victoria 3800, Australia\\
$^{2}$Department of Applied Physics, Polytechnical University of Catalonia, 08860 Barcelona, Spain\\
$^{3}$Institute d’Astronomie et d’Astrophysique, Universit Libre de Bruxelles, ULB CP 226, B-1050 Brussels, Belgium \\
$^{4}$Argelander Institute for Astronomy, University of Bonn, Auf dem Huegel 71, D-53121 Bonn, Germany} 

\begin{document} 
\maketitle

\begin{abstract}
We explore the final fates of massive intermediate-mass stars by computing detailed stellar models from the zero age main sequence until near the end of the thermally pulsing phase. These super-AGB and massive AGB star models are in the mass range between 5.0 and 10.0\,M$_\odot$ for metallicities spanning the range Z=0.02$-$0.0001.
We probe the mass limits $M_{\rm{up}}$, $M_{\rm{n}}$ and $M_{\rm{mass}}$, the minimum masses for the onset of carbon burning, the formation of a neutron star, and the iron core-collapse supernovae respectively, to constrain the white dwarf/electron-capture supernova boundary. 
We provide a theoretical initial to final mass relation for the massive and ultra-massive white dwarfs and specify the mass range for the occurrence of hybrid CO(Ne) white dwarfs.
We predict electron-capture supernova (EC-SN) rates for lower metallicities which are significantly lower than existing values from parametric studies in the literature.
We conclude the EC-SN channel (for single stars and with the critical assumption being the choice of mass-loss rate) is very narrow in initial mass, at most $\approx$ 0.2\,M$_\odot$. This implies that between $\sim$ 2$-$5 per cent of all gravitational collapse supernova are EC-SNe in the metallicity range Z=0.02 to 0.0001. With our choice for mass-loss prescription and computed core growth rates we find, within our metallicity range, that CO cores cannot grow sufficiently massive to undergo a Type 1.5 SN explosion.

\end{abstract}

\section{Introduction\label{sec:Introduction}}

Understanding the evolution, nucleosynthesis and final fates of stars in the mass range 5.0 to 10.0\,M$_\odot$ is important because they bridge the divide between low-mass and high-mass stars, and are relatively numerous, more so than all stars with initial masses greater than 10.0\,M$_\odot$. 
Within this mass range reside super-AGB stars (with masses between $\approx$ 6.5$-$10.0\,M$_\odot$) whose evolution is characterized by off-centre core carbon ignition prior to a thermally pulsing super-AGB (TP-(S)AGB) phase. Owing to their massive cores, they undergo from tens to possibly thousands of thermal pulses. 
The pioneering works by \citet{pap1}, \cite{pap2}, \cite{pap3}, \cite{pap4}, \cite{pap5} explored solar compositions single star\footnote{The binary star evolution channel for this mass range has also been investigated \protect{\cite[e.g.][]{dom93,gil02}}} super-AGB models. Different metallicities and larger grids of super-AGB models have started to populate the literature more recently \cite[e.g.][]{sie06,gil07,poe08,doh10,sie10,ven10a,lug12b,kar12,her12,gil13,ven13,jon13}.

Stars within our mass range cross various critical mass limits, delineating different evolutionary fates. These include; $M_{\rm{up}}$, the minimum mass required to ignite carbon, $M_{\rm{n}}$, the minimum mass for creation of a neutron star, and $M_{\rm{mass}}$, the mass above which stars undergo all stages of nuclear burning and explode as core collapse supernova (CC-SN). Stars with mass between $M_{\rm{up}}$ and $M_{\rm{n}}$ leave ONe white dwarf (WD) remnants whilst stars with masses between $M_{\rm{n}}$ and $M_{\rm{mass}}$ undergo an electron-capture supernova.

Electron capture supernovae (EC-SNe) occur in ONe cores and are caused by the removal of pressure support due to successive electron captures by \chem{24}Mg and \chem{20}Ne in stars with core mass $\sim$ 1.375\,M$_\odot$ \cite[]{miy80,hil84,nom87}. The subsequent rapid contraction leads to oxygen burning, a supernova explosion and the formation of a neutron star. 

There are two (single) star channels that result in an EC-SN. The first channel involves second dredge-up or a dredge-out event reducing the core mass below the Chandrasekhar mass $M_{\rm{CH}}$ and then the evolution through a short thermally pulsing phase to $M_{\rm{CH}}$ \cite[e.g.][]{pap4}. The second channel involves a direct collapse, which occurs when a dredge-out event reduces the core mass close to $M_{\rm{CH}}$ \citep{eld04b}, with these objects designated as ``failed massive stars'' \citep{jon13}. 

As revealed by parametric studies \cite[e.g][]{sie07,poe08}, the main uncertainties limiting our understanding of the fate of super-AGB stars are our poor knowledge of convection, mass-loss rate and the occurrence and efficiency of third dredge-up (3DU). As a consequence, the initial mass range for EC-SNe varies between practically zero to about 1.5\,M$_\odot$ depending on the adopted physics.

The binary fraction of stars increases with increasing mass \citep{zin07}, up to $\sim$70 per cent in O type stars \citep{san12}. Therefore, we may expect stars within our mass range to have a high binary frequency, with this binarity possibly resulting in other EC-SN production channels \citep{pod04}.

At lower metallicities, explosions of CO white dwarfs (Type 1.5 SNe) have been suggested as possible outcomes due to very weak mass-loss \cite[e.g.][]{ibe83,zij04,gil07,lau08,woo11b}. This would place $M_{\rm{n}}$ the neutron star boundary below the carbon burning limit $M_{\rm{up}}$. The very existence of the Type 1.5 SNe channel is highly uncertain, reliant primarily upon the choice of mass-loss rate.

In general, defining the lowest mass for gravitational collapse supernovae is of considerable interest \cite[e.g.][]{heg03,eld04b,ibe13} and the numerous problems which hamper the accurate determination of this boundary, both theoretically and observationally, are still being addressed.  The review by \cite{sma09a} suggests this transition occurs around 8.0 $\pm$ 1\,M$_\odot$ noting this value is highly dependent on stellar model parameters, and in particular on the accurate determination of the core mass at the cessation of central helium burning. 

The white dwarf/gravitational collapse supernova boundary is also important from a galactic chemical evolution perspective due both to the substantial difference in nucleosynthetic products from intermediate-mass compared to massive stars \cite[e.g][]{rom10a,kob11}, and to the differing contribution/feedback from supernova injecting their energy into the environment.
From the point of view of nucleosynthesis, stars of initial masses $8.0-10.0$\,M$_\odot$ have been suggested as the site for r-process production \citep{ish99,ish04}. This led to considerable study of this mass range, in particular to detailed modelling of EC-SNe \cite[e.g][]{nin07,wan03,wan09,qia03,qia07}. Although it now seems unlikely that EC-SN are the main r-process component production site \cite[e.g][]{hof08} the frequency of EC-SNe is still of interest as they may contribute to the weak r-process component \citep{han12} as well as be important contributors to the galactic inventory of a range of isotopes such as \chem{48}Ca and \chem{60}Fe \citep{wan11,wan13a,wan13b}.  

The frequency of EC-SN may help to explain the variety of subclasses of supernovae that have been suggested as arising from EC-SN progenitors. For example the ``supernova impostors'' of SN2008ha, SN2008S and NGC300-OT  \citep{fol09,val09}, and the crab nebula progenitor \citep{dav82,nom82,smi13,tom13,mor14}. 

Recently, detailed evolutionary calculations of super-AGB stars at solar metallicity have been evolved until conditions of EC-SNe \citep{jon13,tak13}.

This paper is the fourth in this series on super-AGB and massive AGB stars. Earlier works have made detailed comparisons between different code results (Paper I - \citealt{doh10}) and explored nucleosynthetic yields of solar, Small and Large Magellanic cloud metallicities (Paper II - \citealt{doh14a}a) and metal-poor and very metal-poor stars (Paper III - \citealt{doh14b}b). 

This work is set out as follows: Section 2 describes our stellar evolution program, Section 3 focuses on the evolution prior to the TP-(S)AGB phase, Section 4 details the thermally pulsing characteristics of these models, Section 5 explores the final fates and compares our results to previous studies in the literature. Finally in Section 6 we present the main conclusions from our work.

\section{Stellar Evolution Program }\label{sec-ev}

The stellar models presented in this work were produced using the Monash stellar evolution program \textsc{monstar} (for details see \citealt{doh10} and references therein.) 
The \textsc{monstar} program includes only seven species: H, \chem{3}He, \chem{4}He, \chem{12}C, \chem{14}N, \chem{16}O and Z$_{\rm{other}}$. Our implementation of carbon burning in this limited network is discussed in detail in Paper I in this series. 
Convection boundaries are determined by using the convective neutrality approach \citep{lat86}. 
Our preferred mass-loss rate is that of \cite{rei75} with $\eta$=1 on the red giant branch following \cite{blo95b} then changing to the \cite{vas93} (their equation 2 - see also \citealt{woo90a}) rate for the carbon burning and thermally pulsing phase. The mass loss rate for the carbon burning phase is highly uncertain and our choice is arbitrary. However, unless a very rapid mass loss prescription is used only a very small amount of material is lost during this phase, given its relatively short duration. We note the \cite{vas93} rate was not derived for super-AGB stars, but for slightly lower mass AGB stars in the initial mass range 0.89$-$5.0 M$_\odot$. No metallicity scaling is applied to the mass-loss rate.

Our initial composition is taken from \cite{gre96} with scaled solar composition. The mixing-length parameter $\alpha_{\rm{mlt}}$ is equal to 1.75 and has been calibrated to the current solar luminosity and radius assuming the age of 4.57 Gyr, (Z/X)$_\odot$ $\sim$ 0.028, and with no overshooting on convective boundaries. There is no clear evidence for the appropriateness of using the solar calibrated value for other phases of evolution or metallicity \cite[e.g.][]{sac91,lyd93,chi95}, but we note that the value of $\alpha_{\rm{mlt}}$ affects the efficiency of convection and hence the stellar evolution. Neutrino losses are from \cite{ito96}. 

Opacities are from OPAL \citep{opal} and for the majority of models we use the compilation of \cite{fer05b} for the low temperature regime. A subset of our models (those of metallicities Z=0.001 and 0.0001) were re-computed (in comparison to models presented in Paper III) to take into account variable composition low temperature opacities from \cite{led09}. The contribution of electron conduction to the total opacity is taken into account using \cite{ibe75a}, \cite{hub69}, \cite{ito83}, \cite{mit84} and \cite{rai82}.
In these calculations we use the equation of state from: \cite{bea71} for relativistic or electron-degenerate gas, the perfect gas equation for fully-ionised regions, and the Saha equation following the method of \cite{bae65} in partially ionised regions.

The calculation of the TP-(S)AGB requires very fine temporal and spatial resolution with the most massive super-AGB star models having in excess of 20 million evolutionary time steps and between 3000$-$5000 spatial mass shells. In order to model the intricate phase of evolution near the end of carbon burning a time-dependent mixing scheme was required and a diffusion approximation was used, as described in \cite{cam08} and \cite{gil13}.  

Detailed models were calculated in the mass range $\sim$ $5.0-10.0$\,M$_\odot$ for metallicities Z=0.02, 0.008, 0.004, 0.001 and 0.0001. Generally divisions of 0.5\,M$_\odot$ in initial mass were used, except near transition masses where a finer initial mass resolution of 0.1\,M$_\odot$ divisions was adopted. A subset of these models served for detailed nucleosynthesis calculations in Papers II and III in this series.

\section{Evolution prior to the Thermally pulsing phase}\label{sec-Super AGB}

The study of intermediate-mass star evolution during the main central burning stages using the \textsc{monstar} program has been described in detail in Paper I in this series. The main difference between that work and the present one is the treatment of the convective boundaries. Previously we used the strict Schwarzschild criterion for convection whilst here (and in Papers II and III) it is replaced by the more physically motivated search for convective neutrality described in \cite{lat86}. The present treatment of convective boundaries results in more massive cores after core helium burning (CHeB) and its effects are similar to those of traditional overshooting applied during this phase. It also leads to a higher oxygen production as additional helium can be mixed in the core and processed by the reaction \chem{12}C($\alpha$,$\gamma$)\chem{16}O which becomes dominant near the end of CHeB. Compared to the classical Schwarzschild approach, we report a $\sim$ 30 per cent increase in the mass of the helium-exhausted core, $\sim$ 30 per cent decrease in the \chem{12}C content and a core helium burning phase about 20 per cent longer. The duration of core H ($\tau_{\rm{HB}}$) and He ($\tau_{\rm{HeB}}$) burning phases, as well as the maximum mass of the convective core during central He burning $M_{\rm{HeB}}$ are provided in Tables~\ref{main} and \ref{main2}. We note that these quantities, in particular $\tau_{\rm{HeB}}$ and $M_{\rm{HeB}}$ are particularly uncertain, due to both the physical (e.g. overshooting (\citealt{ber85}), semi-convection (\citealt{rob72,str03})) and numerical (e.g. numerical diffusion (\citealt{sie07})) sensitivity of modelling the core helium burning phase. 
Prior to the thermally pulsing phase, between 0.1 to 0.5\,M$_\odot$ of mass is lost, with typically greater losses in the more metal rich/and massive models (the total stellar mass at the first thermal pulses, $M_{\rm{Tot}}^{\rm{1TP}}$ can be found in Tables~\ref{main} and \ref{main2}). 

\subsection{Carbon and Advanced Burning}\label{advanceburning}

Degenerate carbon burning operates via a multi-step process consisting of a primary flash (and associated convective region), a convective flame that propagates towards the centre, then subsequent secondary carbon flashes in the outer edges of the CO core. 

The lowest mass stars that ignite carbon do so in the outermost layers of the core then abort further carbon burning (Fig.~\ref{fig-carbon}), leading to the creation of a class of hybrid WDs that we refer to as CO(Ne)s. In these stars the CO core is surrounded by a shell of carbon burnt material $\sim$ 0.1$-$0.4\,M$_\odot$ thick. In \cite{doh10} we examined these aborted carbon ignition flash models with fine resolution and found them not to be an artifact of insufficient spatial zoning. These CO(Ne) cores are also found in a variety of other stellar evolution codes such as \textsc{aton} \citep{ven11a}, \textsc{kepler} (Heger. private communication) and \textsc{mesa} \citep{den13a,che14}, with these later works finding the initial mass width of the CO(Ne) channel dependent on the convective boundary mixing \citep{den13a} and carbon burning reaction rates \citep{che14}.

Here we (re)define\footnote{This is contrary to our definition in Paper I in our series which defined the CO(Ne) cores as not super-AGB stars.}  $M_{\rm{up}}$ as the minimum mass to ignite carbon in which a convective carbon burning region forms. With this definition stars with CO(Ne) cores are more massive than $M_{\rm{up}}$ and hence are called super-AGB stars. Whilst carbon does burn in stars between  $M_{\rm{up}}-0.3\,M_\odot$$\la$ $M_{\rm{ini}}<M_{\rm{up}}$, it does so only partially under radiative conditions and with nuclear luminosities  $L_{\rm{C}}$ $\sim$ 10$-$1000 L$_\odot$, which is relatively energetically modest compared to $L_{\rm{C}}$ $\ga$ $10^6$ L$_\odot$ characteristic of stars of initial masses above M$_{\rm{up}}$.

Analogously to carbon burning in super-AGB stars, models with very massive cores $M_{\rm{C}}$ $\ga$ 1.35\,M$_\odot$ ignite neon off-center, under degenerate conditions with associated convective flashes \citep{pap5,eld04b}. With our limited nuclear network of only seven species we cannot follow nuclear burning stages past carbon burning. We consequently cease calculations when our models reach the approximate temperature for neon ignition $\sim$ 1.2 GK. This condition occur in stars close to the $M_{\rm{mass}}$ limit ($M_{\rm{mass}}$$-0.2$ $M_\odot$ $\la$ $M_{\rm{ini}}<M_{\rm{mass}}$), with the initial masses of these models denoted in bold in Tables~\ref{main} and \ref{main2}. Our calculations were ceased prior to any potential dredge-out event; accordingly we only provide the ONe core mass to this point in Tables~\ref{main} and \ref{main2} in square brackets in the $M_{\rm{2DU}}$ column.

Using a naming convention similar to that for super-AGB stars, models which ignite neon off-center and later reach the thermally pulsing phase, are designated ``hyper''-AGB stars, irrespective of whether this Ne burning ever reaches the centre. With this definition we can more clearly differentiate between the (now) three single star EC-SN channels of: super-AGB star, hyper-AGB star and failed massive star.\footnote{We note this is not the definition of hyper-AGB star used in \protect\cite{eld05}. Whilst we do not, at this stage, expect any substantial difference in either stellar yield, nor energetics between super-AGB and hyper-AGB stars, we feel our definition will prove useful for comparisons between different studies in this very intricate (and highly code dependent) phase of evolution.}

For models at the upper-most end of our mass range, if the ONe core mass exceeds 1.375\,M$_\odot$ at the end of carbon burning, we follow the definition of \cite{sie07}, and class these as massive stars with the lowest initial mass at which this occurs denoted by $M_{\rm{mass}}$ (Table~\ref{tab-crit}). 

\subsection{Second Dredge-up and Dredge-out}\label{sub-2du}

Two main processes can reduce the core mass\footnote{We define the core hereafter as the hydrogen-exhausted core.} prior to the thermally pulsing phase, these being second dredge-up (2DU) and dredge-out. The occurrence of these events is crucial because they will largely dictate the subsequent evolution and fate of the star. If they are not efficient enough and the core mass does not reduce to/or below $M_{\rm{CH}}$, the star proceeds through all nuclear burning phases and will be classified as a massive star.

\begin{figure}
\begin{center}
\includegraphics[width=8cm,angle=0]{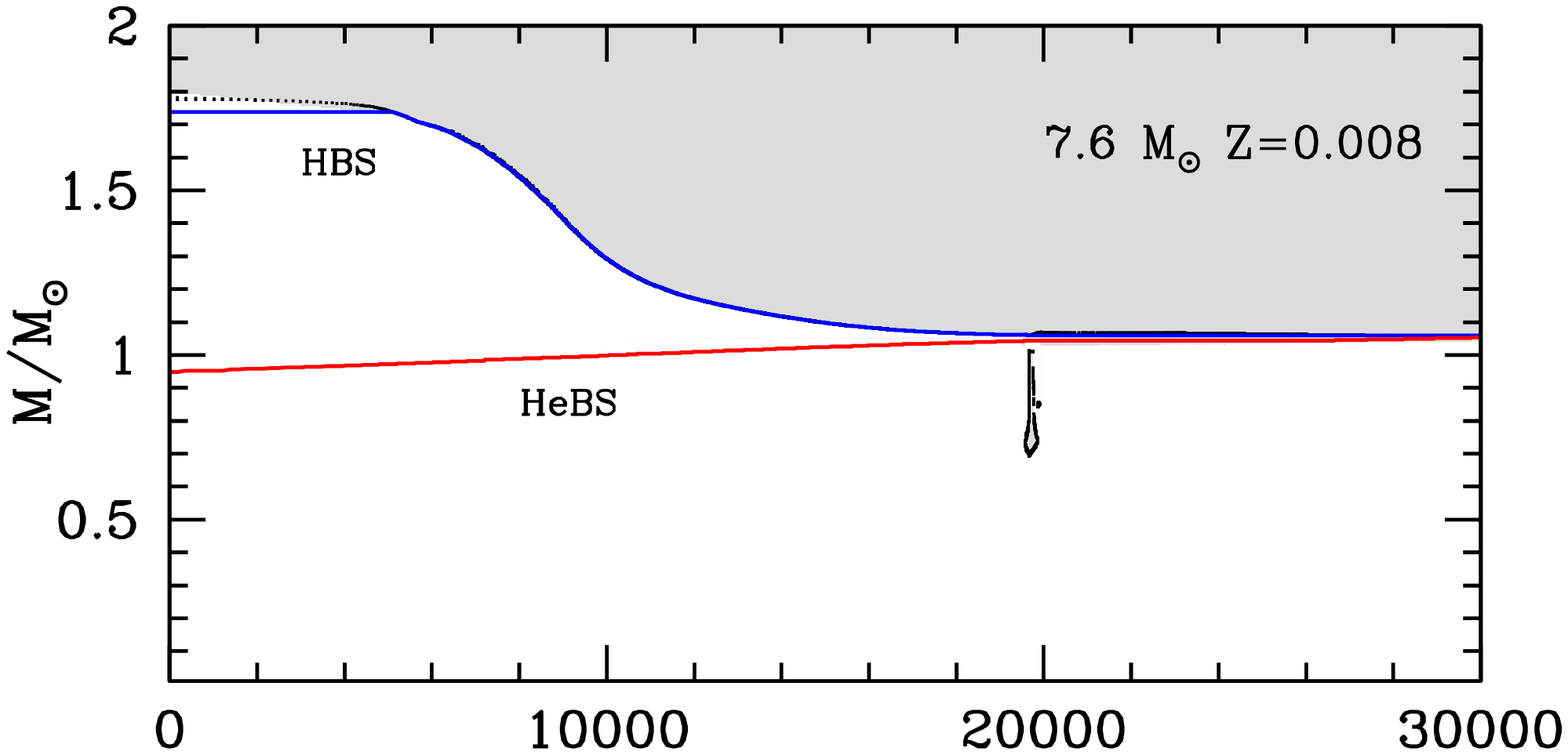} 
\includegraphics[width=8cm,angle=0]{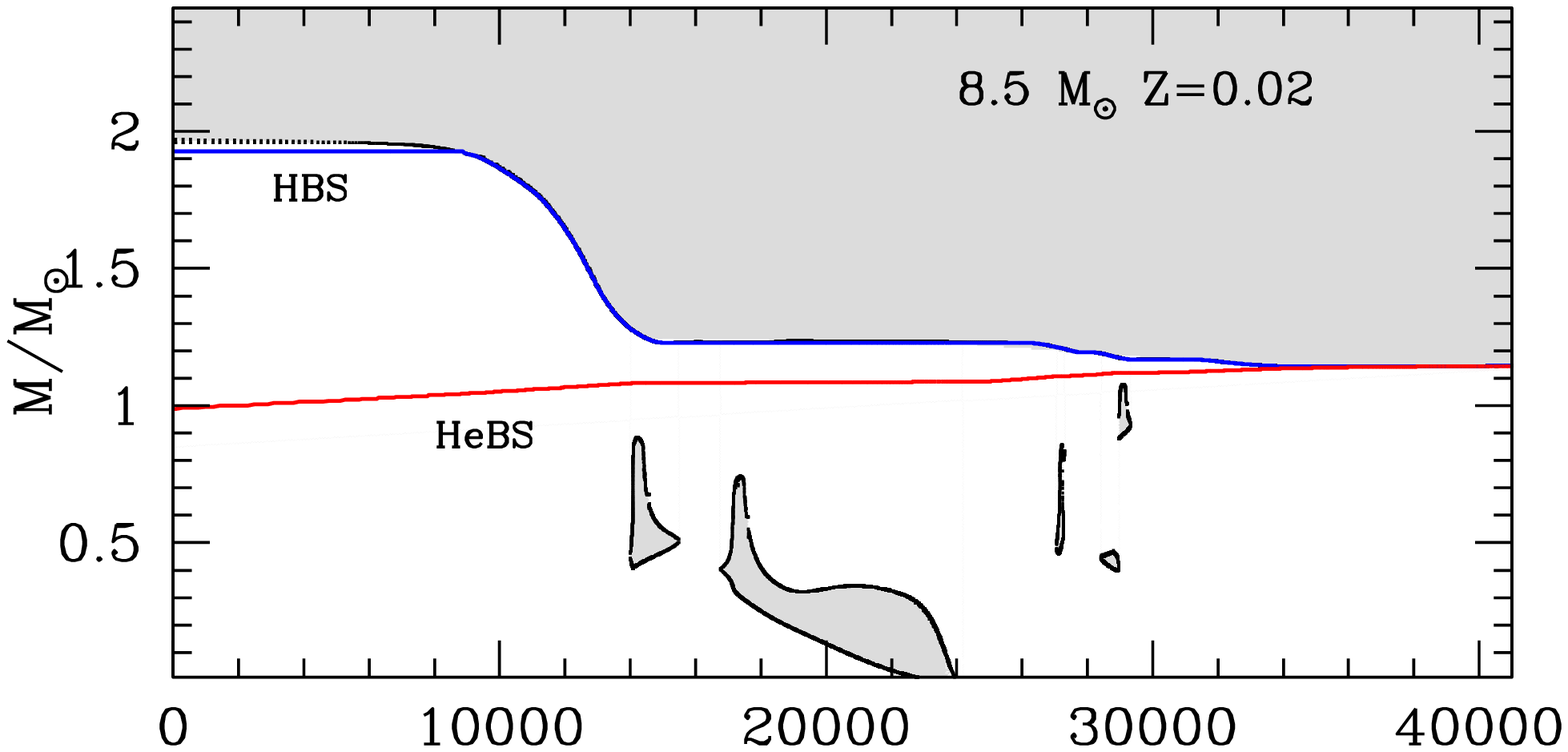}
\includegraphics[width=8cm,angle=0]{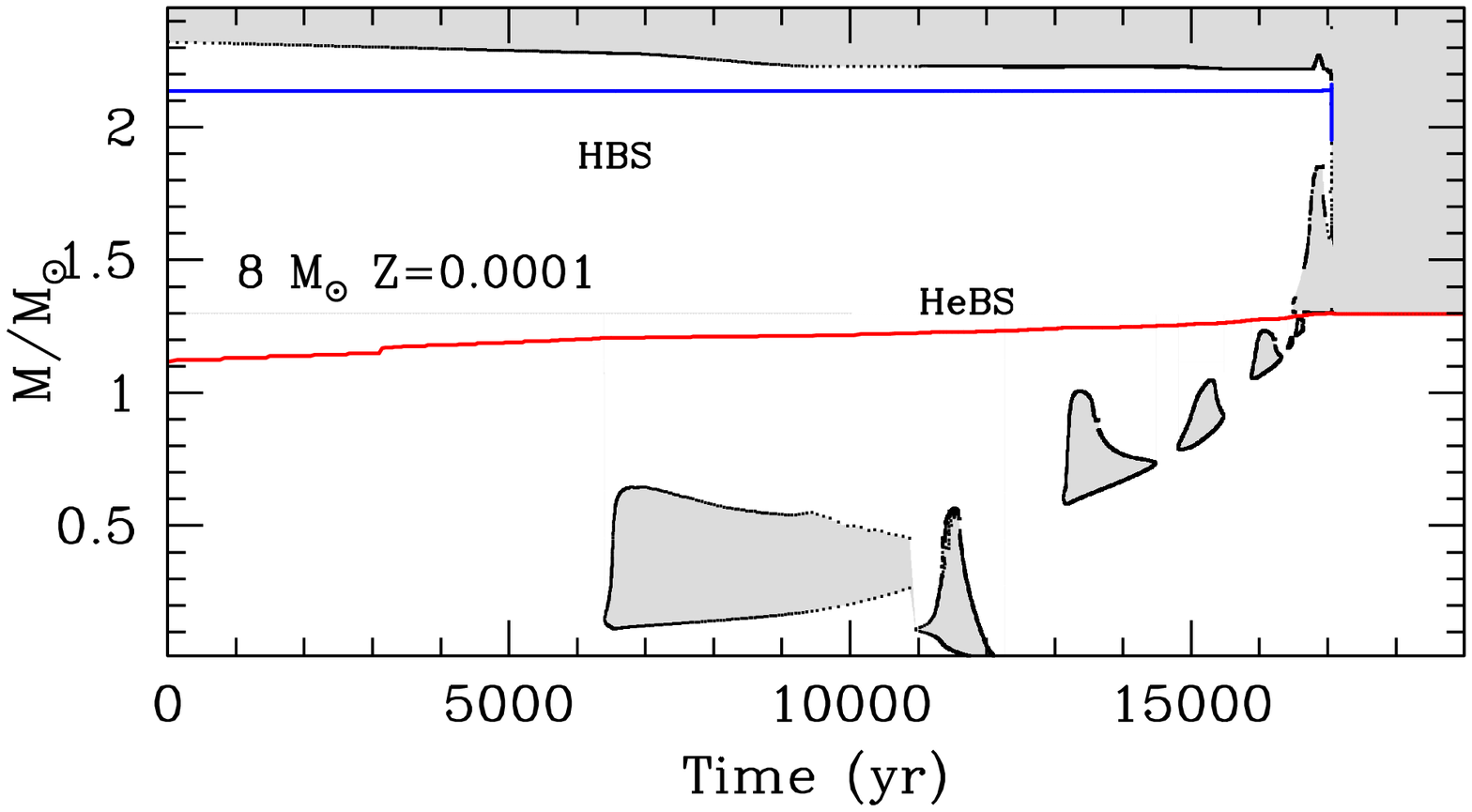}
\caption{Kippenhahn diagrams during the carbon burning phase showing, from top to bottom; an aborted carbon ignition model (7.6\,M$_\odot$ Z=0.008), a model with standard carbon burning and second dredge-up (8.5\,M$_\odot$ Z=0.02) and a model that undergoes dredge-out (8.0\,M$_\odot$ Z=0.0001). The time axes have been offset with the zero when the carbon burning luminosity $L_{\rm{C}}$ exceeds 1\,L$_\odot$. The grey shaded regions represent convection, with those below the helium burning shell showing carbon burning convective regions. The upper part of the convective envelope is omitted from these figures.}
\label{fig-carbon}
\end{center}
\end{figure}

\begin{figure}
\begin{center}
\includegraphics[width=8cm,angle=0]{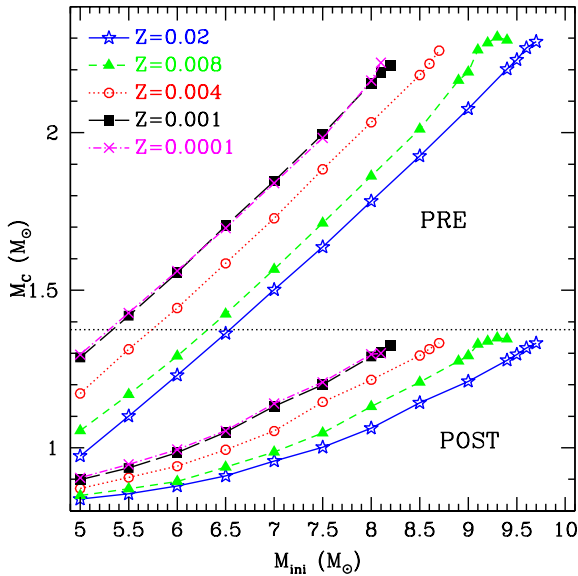}
\caption{Pre- and post- second dredge-up hydrogen-exhausted core masses (M$_\odot$). The dotted horizontal line at 1.375\,M$_\odot$ represents the Chandrasekhar mass $M_{\rm{CH}}$.}
\label{fig-post2du}
\end{center}
\end{figure}

In comparison to intermediate-mass and massive AGB stars in which standard 2DU occurs \citep{bec79}, with increasing initial mass the inward moving convective envelope penetrates very deep, through the extinguished hydrogen burning shell, the intershell region and well into the top of the broad helium burning shell. Whilst this deep dredge-up is not a newly described phenomenon \cite[e.g][]{bec79,her04b,sie06,lau07,sud10} we have proposed a new nomenclature of ``corrosive 2DU'' \citep{gil13,doh14b}. In our models, corrosive 2DU occurs in stars with core masses $M_{\rm{C}}$ $\approx$ 1.15\,M$_\odot$$-$1.28\,M$_\odot$, with this value lower in the lower metallicity models, primarily due to their broader He shells. 

The most massive super-AGB star models undergo a dredge-out event \citep{pap5,sie07}. This type of event occurs near the end of the carbon burning phase when the massive ONe core is contracting, and releases a large amount of gravitational energy, which heats up the He-rich layers. The temperature increase induces the development of a convective instability in the helium shell, which widens, and eventually meets, the inward moving convective envelope. When these two convective zones merge, fresh protons from the convective envelope are mixed downwards to a higher temperature region, where they are rapidly burnt in a violent episode resulting in a hydrogen flash. This proton ingestion episode is similar in nature to those described in extremely metal deficient stars \cite[e.g][]{fuj90,chi01,sie02,cam08,lau09}. In this process there is formation of a \chem{13}C rich region, a substantial release of neutrons via the \chem{13}C($\alpha$,n)\chem{16}O reaction with production of heavy elements. This is similar to the neutron super-burst described in \cite{cam10}. We explore heavy element production in dredge-out events in a forthcoming work.

Very fine temporal and spatial resolution near the end of carbon burning phase is crucial in these massive super AGB models to accurately follow the intensity and duration of the secondary carbon flashes (which can be modified by changes in resolution). These flashes are relevant as they provide the necessary conditions (the interplay between the gravitational and nuclear burning luminosities) to cause the dredge out event. In addition to resolution, these convective behaviours are very dependent on the mixing processes.
We also note that in our models, after a dredge-out event there is a very short period in time when both the H and He shells are fully extinguished. During this time, there is a slight penetration of the convective envelope into the CO core. This further reduces the core mass, which could later impact the star's future fate and the frequency, or even ultimately the occurrence, of the EC-SN channel.

In models with masses slightly less than those which undergo a dredge-out event, a helium burning convective region also develops near the end of the carbon burning phase. However, it remains isolated from the H-rich convective envelope. Although this partially burnt region with its abundant carbon will later be consumed by the inward moving envelope, the occurrence of this type of event could in principle be observationally distinguishable from a dredge-out by the lack of neutron-rich material in the stellar envelope.

In Fig.~\ref{fig-carbon} we provide Kippenhahn diagrams for a variety of cases during the carbon burning and dredge-up phases. They highlight the different behaviours described above, such as: aborted carbon ignition (top panel), standard carbon burning with second dredge-up (middle panel) and a dredge-out event (bottom panel).

We note that the boundaries in initial mass between standard 2DU, corrosive 2DU and dredge-out events are highly model dependent. All these mixing events apart from standard 2DU enrich the surface with large amounts of \chem{12}C, and, in the majority of cases, lead to the formation of carbon stars. This surface enrichment has important consequences related to mass loss which we discuss in section~\ref{subsec-ml}.

In Fig.~\ref{fig-post2du} we show the pre- and post- 2DU/dredge-out core mass as a function of the initial mass in the metallicity range of our models. Whilst the pre- 2DU/dredge-out core masses show a clear linear trend with increasing initial mass for a given metallicity, this is not the case for the post- 2DU/dredge-out core masses. There is a flatter slope in the post- 2DU/dredge-out core masses versus mass relation for the lower core masses. This is apparently coincident with the mass at which the core composition changes from CO to ONe ($\approx$ 1.06\,M$_\odot$). We attribute this change in slope to the stalling of the inward moving convective envelope due to core carbon ignition, and the subsequently longer duration for the helium burning shell to grow outwards in mass during the carbon burning phase. 

The post- 2DU/dredge-out values can be found in Tables~\ref{main} and \ref{main2}. The similarity in behaviour for the Z=0.001 and 0.0001 models is a consequence of very similar composition, particularly in the CNO content.

\section{Thermally Pulsing Super-AGB and Massive AGB Evolution}\label{sec-tp}

\begin{figure}
\begin{center}
\resizebox{\hsize}{!}{\includegraphics{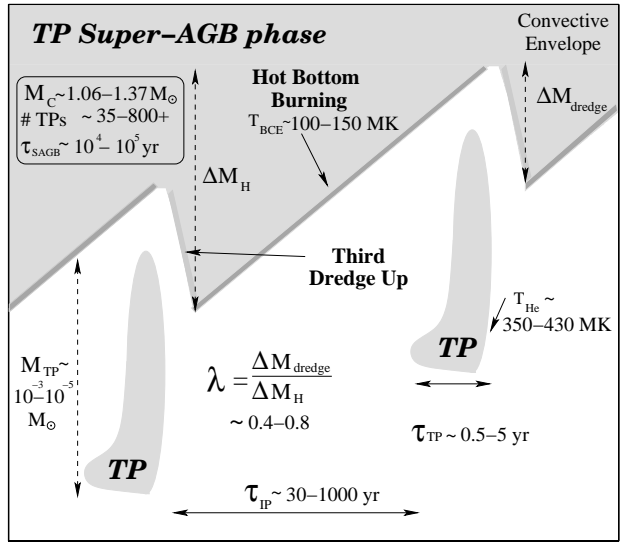}}
\caption{Schematic Kippenhahn diagram of two consecutive thermal pulses showing typical values for super-AGB stars. The light grey shaded regions represent convective regions. Variables are; the duration and mass involved in the TP convective pocket $\tau_{\rm{TP}}$ and $M_{\rm{TP}}$ respectively; $\Delta$$M_{\rm{H}}$ is the increase in the core mass during the interpulse period; $\Delta$$M_{\rm{dredge}}$ is the depth of dredge-up, $T_{\rm{BCE}}$ is the temperature at the base of the convective envelope and $T_{\rm{He}}$ is the maximum temperature in the helium burning intershell region. For details of further variables refer to Table~\ref{main}.}
\label{fig-schematic}
\end{center}
\end{figure}

In this section we describe the most relevant TP-(S)AGB characteristics, with special attention to those that affect more directly the final outcome of our models. A selection of these parameters is provided in Tables~\ref{main} and \ref{main2}. Fig.~\ref{fig-schematic} gives a schematic overview of the typical values associated with the thermally pulsing phase of super-AGB stars from our models.

After completion of either 2DU or dredge-out, the core contracts and both the hydrogen burning shell and helium burning shells are (re)established, leading to the development of recurrent thermal instabilities, with the frequency of these events increasing with increasing core mass. 

The competition between the core growth and the mass-loss will determine the final fate of these objects. If the core mass $M_{\rm{C}}$ reaches the Chandrasekhar mass $M_{\rm{CH}}$ of $\approx$ 1.375\,M$_\odot$ \citep{nom84} before the envelope is lost, the star will undergo an electron-capture supernova.

In the following subsections we explore in further detail these competing processes and the factors which affect them. We then describe an interesting phenomenon which also affects the final evolution of super-AGB and massive AGB stars: the possible envelope ejection due to the iron peak opacity instability \citep{swe99,lau12}.

\subsection{Core Growth Rate}\label{subcgr}

The \textit{effective} core growth rate depends on the actual core growth rate due to the outwardly moving hydrogen burning shell and on the third dredge-up which acts repeatedly to reduce the core mass after each thermal pulse.

The core growth is driven by the hydrogen burning shell, which reaches temperatures in the range $\sim$ 80$-$150MK, and is metallicity dependent with lower metallicity models achieving higher temperatures.

According to our calculations, the average core growth rates for the entire TP-(S)AGB phase $\langle$$\dot{M}_{\rm{C}}$$\rangle$, which is the average of the contribution from each individual TP\footnote{Defined as $\dot{M}_{\rm{C}}$=$\Delta$M$_{\rm{H}}$/$\tau_{\rm{IP}}$ where $\Delta$M$_{\rm{H}}$ is the increase in the core mass during the interpulse period and $\tau_{\rm{IP}}$ is the interpulse period.}, are given in Tables~\ref{main} and \ref{main2}. The $\langle$$\dot{M}_{\rm{C}}$$\rangle$ values range from 4$-$8$\times$10$^{-7}$\,M$_\odot$ yr$^{-1}$, with faster rates in the more massive and/or metal-rich stars. We note that these values do \textit{not} take into account 3DU. These values agree well with those from previous works \cite[e.g][]{nom84,poe07,sie10}. A strong anti-correlation between core growth rate and core radius was noted by \cite{sie10}, with lower metallicity models having slower core growth rates by virtue of their more condensed structure. 

The efficiency of 3DU is one of the most important, yet poorly constrained, aspects of AGB modelling, especially at the larger core masses included in this work. In computations of super-AGB stars the efficiency of 3DU ranges from: no 3DU ($\lambda$=0) as found by \cite{sie10} and \cite{ven13}; slight 3DU ($\lambda\sim0.07-0.30$) found by \cite{pap2}; efficient 3DU ($\lambda\sim0.4-0.8$) found in this work and \cite{gil13}; or extreme 3DU ($\lambda$$\approx$1) found by \cite{her12}. This large variation in 3DU efficiency may lead to a large disparity between final fate calculations. 
In this study we find average $\langle$$\lambda$$\rangle$ values between $\sim$ 0.4$-$0.8, as reported in Tables~\ref{main} and \ref{main2}. The maximum efficiency $\lambda_{\rm{Max}}$ values are between $\sim$ 0.5 and 0.95, with the efficiency of 3DU found to decrease with increasing initial mass.

Whilst we defer further discussion of uncertainties in 3DU occurrence/efficiency to our forthcoming work (Paper V - Doherty et. al in preparation), we do wish to highlight the nucleosynthetic evidence for 3DU in massive AGB stars. Large over-abundances of Rb have been observed in Magellanic cloud and Galactic O-rich AGB stars with initial masses $\sim$ 4$-$8\,M$_\odot$ \cite[]{gar06,gar09}. This Rb is suspected to be produced in high neutron density environments where the \chem{22}Ne neutron source is activated. High temperatures are required to activate this neutron source, with the necessary conditions only found at the base of the thermal pulse in AGB stars with initial masses $M_{\rm{ini}}$ $\ga$ 4\,M$_\odot$ \citep{ibe77,abi01,van12}.

\begin{table*}
\caption{Model properties: $M_{\rm{ini}}$ is the initial mass; $\tau_{\rm{HB}}$ is the main sequence duration; $M_{\rm{HeB}}$ is the maximum mass of the convective core during helium burning; $\tau_{\rm{HeB}}$ is the duration of the core helium burning phase; X$_{\rm{C}}$ is the central carbon mass fraction at the cessation of core He burning; $M_{\rm{2DU}}$ is the maximum extent of 2DU (defined by depth of the inner edge of the convective envelope); $M_{\rm{Tot}}^{\rm{1TP}}$ is the stellar mass at the first thermal pulse; $\langle$$\lambda$$\rangle$ is the average third dredge-up parameter; $\langle$$\tau_{\rm{IP}}$$\rangle$ is the average interpulse period, $L_{\rm{MAX}}$ is the maximum quiescent luminosity during the interpulse phase, $\langle$$\dot{M}_{\rm{C}}$$\rangle$ is the average core growth rate during the TP-(S)AGB phase; $\langle$$\dot{M}$$\rangle$ is the average mass-loss rate during the TP-(S)AGB phase; $M_{\rm{C}}^{F}$ and $M_{\rm{env}}^{F}$ are the core and envelope masses of the last computed model; $M_{\rm{C}}^{Ex}$ is the extrapolated core mass (see text for extrapolation details); $\tau_{\rm{(S)AGB}}$ is the duration of the thermally pulsing phase; N$_{\rm{TP}}$ is the number of thermal pulses; $\tau_{\rm{Total}}$ is the total stellar lifetime whilst the last column identifies the resultant stellar remnant. All variables are tabulated until the cessation of detailed computed models. An initial mass in bold represents models in which the conditions for neon ignition were reached. These stars are likely to end their lives as neutron stars after undergoing an EC-SN. Models with an asterisk in the last column may end their lives as EC-SN, depending on the particular choice of extrapolation inputs. Values in the $M_{\rm{2DU}}$ column in square brackets represent models which ceased prior to dredge-out, with this tabulated value the ONe core mass. Note that $n(m)= n\times 10^{m}$.}
\begin{center}\setlength{\tabcolsep}{1.5pt} 
\begin{tabular}{ccccccccccccccccccr}
\hline 
$M_{\rm{ini}}$&$\tau_{\rm{HB}}$&$M_{\rm{HeB}}$&$\tau_{\rm{HeB}}$&X$_{\rm{C}}$&$M_{\rm{2DU}}$&$M_{\rm{Tot}}^{\rm{1TP}}$&$\langle$$\lambda$$\rangle$&$\langle$$\tau_{\rm{IP}}$$\rangle$&$L_{\rm{MAX}}$&$\langle$$\dot{M}_{\rm{C}}$$\rangle$&$\langle$$\dot{M}$$\rangle$&$M_{\rm{C}}^{F}$&$M_{\rm{env}}^{F}$&$M_{\rm{C}}^{Ex}$&$\tau_{\rm{(S)AGB}}$&N$_{\rm{TP}}$&$\tau_{\rm{Total}}$ &WD/NS \\ 
(M$_\odot$)&(Myr)&(M$_\odot$)&(Myr)&&(M$_\odot$)&(M$_\odot$)&&(yr)&(L$_\odot$)&(M$_\odot$/yr)&(M$_\odot$/yr)&(M$_\odot$)&(M$_\odot$)&(M$_\odot$)&(yr)&&(Myr) & Remnant \\
\hline  \multicolumn{19}{c}{Z=0.02} \\ \hline 
5.0 &84.85&0.457&24.49&0.28&0.833&4.91&0.78&1.04(4)&2.76(4)&3.26(-7)&1.01(-5)&0.852&1.23&0.860&2.81(5)&28&113.98 &CO WD \\
5.5 &72.18&0.513&18.40&0.29&0.858&5.40&0.81&8.60(3)&3.01(4)&3.55(-7)&1.61(-5)&0.870&1.33&0.876&2.75(5)&33&89.41 &CO WD\\
6.0 &55.52&0.576&14.43&0.28&0.882&5.88&0.80&6.13(3)&3.40(4)&3.93(-7)&2.82(-5)&0.895&1.42&0.899&2.08(5)&35&72.37 &CO WD\\
6.5 &46.57&0.649&11.67&0.26&0.914&6.36&0.77&4.54(3)&3.86(4)&4.34(-7)&2.62(-5)&0.928&1.32&0.933&1.57(5)&36&60.02 &CO WD \\ 
7.0 &38.65&0.723&9.62 &0.24&0.958&6.85&0.75&3.04(3)&4.38(4)&4.77(-7)&3.91(-5)&0.967&1.54&0.972&1.11(5)&38&50.78 &CO WD \\ 
7.5 &34.50&0.786&7.91 &0.28&1.002&7.34&0.75&2.08(3)&5.00(4)&5.28(-7)&5.14(-5)&1.011&1.63&1.015&9.14(4)&45&43.47 &CO WD\\
8.0 &30.45&0.877&6.73 &0.26&1.062&7.70&0.77&1.28(3)&6.00(4)&5.97(-7)&6.45(-5)&1.072&1.61&1.075&7.78(4)&62&37.93&CO(Ne) WD\\
8.5 &27.01&0.957&5.76 &0.27&1.143&8.02&0.76&6.08(2)&7.33(4)&6.82(-7)&7.71(-5)&1.154&1.78&1.158&6.60(4)&110&33.39&ONe WD\\
9.0 &24.28&1.052&4.98 &0.26&1.211&8.54&0.71&2.89(2)&8.65(4)&7.50(-7)&8.61(-5)&1.225&1.82&1.230&6.39(4)&221&29.81&ONe WD\\
9.4 &22.44&1.086&4.55 &0.22&1.278&9.02&0.60&1.10(2)&9.81(4)&8.08(-7)&1.00(-4)&1.295&2.45&1.303&5.27(4)&479&27.41&ONe WD\\
9.5 &22.02&1.161&4.40 &0.25&1.296&9.14&0.55&7.73(1)&1.02(5)&8.21(-7)&1.04(-4)&1.315&2.55&1.324&5.09(4)&659&26.82&ONe WD\\ 
9.6 &21.61&1.158&4.39 &0.18&1.317&9.27&0.52&5.33(1)&1.05(5)&8.34(-7)&1.12(-4)&1.334&$\it{3.21}$&1.345&4.22(4)&792&26.41& ONe WD \\ 
9.7 &21.23&1.191&4.17 &0.25&1.332&9.37&0.47&3.98(1)&1.11(5)&8.41(-7)&1.29(-4)&1.343&$\it{5.10}$&1.361&2.27(4)&572&25.74&ONe WD \\ 
$\bf{9.8}$&20.85&1.257&4.20 &0.19&[1.356]&-&-&-&-&-&-&-&-&-&-&-&25.35&EC-SN/NS \\
\hline \multicolumn{19}{c}{Z=0.008} \\ \hline
5.0 &78.53&0.517&22.48&0.27&0.851&4.89&0.90&1.01(4)&3.50(4)&3.66(-7)&5.18(-6)&0.869&0.88&0.875&6.06(5)&61&105.57&CO WD \\
5.5 &63.79&0.581&17.16&0.25&0.873&5.37&0.89&7.80(3)&3.87(4)&3.95(-7)&6.91(-6)&0.894&0.92&0.900&5.15(5)&67&84.21&CO WD \\
6.0 &53.09&0.625&13.37&0.30&0.899&5.85&0.88&6.14(3)&4.36(4)&4.21(-7)&9.60(-6)&0.916&1.10&0.922&3.99(5)&66&68.86&CO WD\\
6.5 &45.09&0.690&10.78&0.30&0.939&6.32&0.86&4.18(3)&4.89(4)&4.56(-7)&1.71(-5)&0.951&0.95&0.955&2.58(5)&63&57.66&CO WD \\
7.0 &38.94&0.773& 8.92&0.29&0.986&6.80&0.82&2.71(3)&5.59(4)&4.93(-7)&3.19(-5)&0.996&1.05&0.999&1.49(5)&56&49.16 &CO WD \\
7.5 &34.13&0.869& 7.55&0.26&1.047&7.27&0.79&1.59(3)&6.43(4)&5.47(-7)&5.67(-5)&1.057&1.14&1.059&8.95(4)&58&42.64 &CO WD \\
7.6 &33.29&0.886& 7.31&0.26&1.059&7.37&0.79&1.39(3)&6.66(4)&5.57(-7)&6.29(-5)&1.069&1.23&1.071&8.05(4)&59&41.50 &CO(Ne) WD \\
8.0 &30.27&0.961&6.46 &0.24&1.131&7.56&0.78&7.21(2)&7.66(4)&6.31(-7)&7.90(-5)&1.140&1.21&1.142&6.60(4)&93&37.43 &ONe WD \\
8.5 &27.14&1.072&5.60 &0.21&1.208&7.96&0.73&3.17(2)&9.35(4)&7.01(-7)&9.11(-5)&1.219&1.42&1.222&5.84(4)&185&33.29 &ONe WD\\
8.9 &25.04&1.142&4.92 &0.25&1.275&8.47&0.65&1.24(2)&1.10(5)&7.65(-7)&1.05(-4)&1.290&1.67&1.294&5.26(4)&426&30.47 &ONe WD\\ 
9.0 &24.56&1.165&4.84 &0.23&1.292&8.59&0.58&9.00(1)&1.12(5)&7.69(-7)&1.09(-4)&1.309&1.76&1.314&5.07(4)&565&29.88 &ONe WD\\
9.1 &24.09&1.207&4.72 &0.24&1.329&8.76&0.49&4.32(1)&1.20(5)&8.04(-7)&1.25(-4)&1.342&$\it{3.15}$&1.352&3.40(4)&768&29.26 &ONe WD\\
9.2 &23.65&1.220&4.59 &0.23&1.338&8.87&0.45&3.51(1)&1.23(5)&7.99(-7)&1.43(-4)&1.348&$\it{4.46}$&1.362&2.14(4)&612&28.66&ONe WD\\
9.3 &23.22&1.225&4.47 &0.25&1.349&8.97&0.40&2.64(1)&1.25(5)&8.05(-7)&1.46(-4)&1.359&$\it{4.63}$&1.373&2.03(4)&770&28.05&ONe WD*\\
9.4 &22.80&1.244&4.56 &0.17&1.345&9.01&0.44&3.20(1)&1.24(5)&7.99(-7)&1.52(-4)&1.352&$\it{5.40}$&1.368&1.48(4)&463&27.70&ONe WD* \\
$\bf{9.5}$ &22.40&1.250&4.39& 0.22&[1.360]& - &-  &  -& -&-&-&-&-&-&-&-&27.15&EC-SN/NS \\
 \hline \multicolumn{19}{c}{Z=0.004} \\ \hline
5.0 &74.84&0.560&20.12&0.31&0.869&4.89&0.88&8.13(3)&4.06(4)&3.81(-7)&4.78(-6)&0.898&0.77&0.905&6.75(5)&84&99.00 &CO WD\\
5.5 &61.18&0.658&16.02&0.24&0.908&5.38&0.89&5.54(3)&4.68(4)&4.27(-7)&9.08(-6)&0.925&0.68&0.929&4.15(5)&76&80.02 &CO WD\\
6.0 &51.42&0.728&12.60&0.25&0.948&5.87&0.85&4.07(3)&5.28(4)&4.48(-7)&1.41(-5)&0.966&0.80&0.970&2.92(5)&73&66.03 &CO WD \\
6.5 &43.83&0.777&10.10&0.29&0.994&6.37&0.84&2.60(3)&6.00(4)&5.04(-7)&2.61(-5)&1.005&0.83&1.008&1.73(5)&68&55.46 &CO WD\\
7.0 &38.06&0.852&8.35 &0.28&1.053&6.85&0.80&1.46(3)&6.98(4)&5.49(-7)&5.29(-5)&1.063&0.86&1.065&9.33(4)&65&47.49 &CO WD \\
7.1 &37.07&0.884&8.05 &0.29&1.064&6.95&0.79&1.31(3)&7.15(4)&5.61(-7)&5.66(-5)&1.074&0.98&1.076&8.66(4)&67&46.13 &CO(Ne) WD \\
7.5 &33.53&0.979&7.15 &0.23&1.145&7.26&0.76&5.98(2)&8.28(4)&6.36(-7)&8.27(-5)&1.155&1.03&1.157&6.16(4)&104&41.53&ONe WD\\
8.0 &29.85&1.073&6.10 &0.24&1.216&7.54&0.71&2.90(2)&1.00(5)&7.05(-7)&9.61(-5)&1.227&1.19&1.230&5.35(4)&192&36.59 &ONe WD\\
8.5 &26.86&1.166&5.25 &0.23&1.293&8.13&0.60&9.06(1)&1.20(5)&7.23(-7)&1.13(-4)&1.308&1.41&1.312&4.81(4)&532&32.66 &ONe WD\\
8.6 &26.33&1.185&5.11 &0.24&1.313&8.26&0.55&6.02(1)&1.25(5)&7.75(-7)&1.17(-4)&1.330&1.41&1.334&4.73(4)&787&31.94 &ONe WD\\ 
8.7 &25.85&1.226&5.08 &0.18&1.333&8.38&0.43&3.92(1)&1.28(5)&7.70(-7)&1.34(-4)&1.347&$\it{2.63}$&1.356&3.29(4)&839&31.35 &ONe WD\\ 
$\bf{8.8}$&25.32&1.241&4.86 &0.22&[1.350]&-&-&-&-&-&-&-&-&-&-&-&30.59&EC-SN/NS \\
$\bf{8.9}$&24.85&1.263&4.77 &0.22&[1.369]&-&-&-&-&-&-&-&-&-&-&-&30.00&EC-SN/NS \\
\hline
\multicolumn{19}{c}{Continued on next page} \\  
\hline
\end{tabular}
\end{center}\label{main}
\end{table*}

\begin{table*}
\caption{Continuation of Table~\ref{main} but for metallicities Z=0.001 and 0.0001.}
\begin{center}\setlength{\tabcolsep}{1.5pt} 
\begin{tabular}{ccccccccccccccccccr}
\hline
$M_{\rm{ini}}$&$\tau_{\rm{HB}}$&$M_{\rm{HeB}}$&$\tau_{\rm{HeB}}$&X$_{\rm{C}}$&$M_{\rm{2DU}}$&$M_{\rm{Tot}}^{\rm{1TP}}$&$\langle$$\lambda$$\rangle$&$\langle$$\tau_{\rm{IP}}$$\rangle$&$L_{\rm{MAX}}$&$\langle$$\dot{M}_{\rm{C}}$$\rangle$&$\langle$$\dot{M}$$\rangle$&$M_{\rm{C}}^{F}$&$M_{\rm{env}}^{F}$&$M_{\rm{C}}^{Ex}$&$\tau_{\rm{(S)AGB}}$&N$_{\rm{TP}}$&$\tau_{\rm{Total}}$ &WD/NS \\ 
(M$_\odot$)&(Myr)&(M$_\odot$)&(Myr)&&(M$_\odot$)&(M$_\odot$)&&(yr)&(L$_\odot$)&(M$_\odot$/yr)&(M$_\odot$/yr)&(M$_\odot$)&(M$_\odot$)&(M$_\odot$)&(yr)&&(Myr) & Remnant \\
\hline
\multicolumn{19}{c}{Z=0.001} \\ 
\hline  
5.0 &70.33&0.632&16.68& 0.28&0.899&4.94&0.90&5.78(3)&4.85(4)&4.34(-7)&6.02(-6)&0.918&0.99&0.925&5.03(5)&88 &90.38&CO WD\\
5.5 &58.19&0.696&12.94& 0.30&0.937&5.43&0.93&4.17(3)&5.67(4)&4.67(-7)&1.01(-5)&0.953&0.78&0.956&3.66(5)&89 &73.62 &CO WD\\
6.0 &49.22&0.774&10.43& 0.29 &0.985&5.93&0.85&2.57(3)&6.59(4)&5.08(-7)&1.64(-5)&1.002&0.89&1.006&2.46(5)&97 &61.33 &CO WD\\
6.5 &42.31&0.867&8.67 & 0.26 &1.058&6.43&0.80&1.33(3)&7.59(4)&5.47(-7)&3.75(-5)&1.070&1.15&1.073&1.12(5)&87 &52.18 &CO(Ne) WD \\
7.0 &36.94&0.953&7.18 & 0.27&1.132&6.93&0.76&6.37(2)&8.75(4)&6.18(-7)&7.42(-5)&1.141&0.94&1.143&6.52(4)&104&45.01 &ONe WD \\
7.5 &32.67&1.047&6.12 &0.26 &1.207&7.15&0.68&2.87(2)&1.02(5)&6.83(-7)&1.09(-4)&1.218&1.11&1.220&4.41(4)&167&39.54 &ONe WD \\
8.0 &29.23&1.161&5.32 & 0.24&1.290&7.77&0.51&8.54(1)&1.22(5)&7.38(-7)&1.21(-4)&1.305&1.36&1.309&4.22(4)&494&35.13 &ONe WD \\
8.1 &28.62&1.182&5.22 & 0.22&1.303&7.88&0.55&7.39(1)&1.23(5)&7.39(-7)&1.26(-4)&1.317&1.47&1.321&4.04(4)&556&34.39&ONe WD \\
8.2 &28.03&1.209&5.07 & 0.22&1.324&7.98&0.47&4.75(1)&1.25(5)&7.47(-7)&1.28(-4)&1.340&1.51&1.349&4.01(4)&845&33.63&ONe WD \\
$\bf{8.3}$&27.46&1.214&4.87 &0.26 &[1.351]&-&-&-&-&-&-&-&-&-&-&-&32.81&EC-SN/NS \\
\hline \multicolumn{19}{c}{Z=0.0001} \\ \hline
5.0 &66.96&0.640&14.8 &0.29 &0.905&4.96&0.92&6.14(3)&6.29(4)&4.73(-7)&3.70(-6)&0.921&0.86&0.930&8.60(5)&141&85.58&CO WD\\
5.5&55.75&0.715&11.89 &0.28 &0.947&5.46&0.91&3.47(3)&7.36(4)&5.22(-7)&6.15(-6)&0.972&0.81&0.978&5.95(5)&173&70.35&CO WD\\
6.0&47.35&0.766&9.66  &0.28&0.995&5.96&0.89&2.26(3)&8.38(4)&5.55(-7)&9.84(-6)&1.018&0.89&1.023&4.12(5)&183&58.93&CO WD\\
6.5&40.89&0.841&8.01  &0.28 &1.063&6.46&0.83&1.20(3)&9.89(4)&5.84(-7)&1.81(-5)&1.084&0.77&1.088&2.55(5)&214&50.38&CO(Ne) WD \\
7.0&35.85&0.961&6.82  & 0.26&1.144&6.96&0.78&4.69(2)&1.16(5)&6.56(-7)&4.00(-5)&1.168&0.76&1.171&1.25(5)&269&43.63&ONe WD \\
7.5&31.80&1.038&5.82  & 0.26&1.210&7.43&0.67&2.52(2)&1.32(5)&7.03(-7)&8.83(-5)&1.225&0.70&1.227&6.23(4)&248&38.39& ONe WD \\
8.0&28.51&1.168&5.08  &0.23 &1.297&7.77&0.54&7.73(1)&1.47(5)&7.35(-7)&1.39(-4)&1.310&1.07&1.313&3.88(4)&503&34.18&ONe WD \\
8.1&27.93&1.171&4.90  &0.26 &1.306&7.87&0.50&6.61(1)&1.54(5)&7.43(-7)&1.42(-4)&1.321&1.04&1.324&3.87(4)&626&33.40&ONe WD \\
$\bf{8.2}$&27.37&1.212&4.86 & 0.20 &[1.349]&-&-&-&-&-&-&-&-&-&-&-&32.72&EC-SN/NS \\
$\bf{8.3}$&26.84&1.234&4.72 & 0.21&[1.359]&-&-&-&-&-&-&-&-&-&-&-&31.97&EC-SN/NS \\
\hline 
\end{tabular}
\end{center}\label{main2}
\end{table*}

 \subsection{Mass Loss}\label{subsec-ml}

 Mass loss is fundamental in stellar modeling in general and, in particular, to determine the final fates of stars. Unfortunately it is also one of the largest unknowns, with the main driver of mass-loss in AGB stars, especially at lower metallicities, not yet fully understood. Whilst traditionally the mass-loss rate was thought to be lower at low metallicity \cite[e.g.][]{bow91,wil00,zij04}, recent theoretical \cite[e.g.][]{mat08b} and observational \cite[e.g.][]{gro09,lag08} studies suggest this may not always be the case. The \cite{blo95} mass-loss prescription is also commonly used by low-metallicity AGB star modellers \cite[e.g.][]{her04a,ven13} and with its high luminosity exponent it results in \textit{faster} mass-loss at lower metallicities.

Also important when considering mass-loss is the surface composition. The dredge-up or dredge-out events can enrich the surface composition to metallicities comparable to, or even greater than, solar values. For these reasons we use the empirical mass-loss rate by \cite{vas93} (their equation 2) with no metallicity scaling. For massive AGB stars this leads to a mass-loss rate that is initially quite moderate $\sim$ 10$^{-7}$ M$_\odot$ yr$^{-1}$, until a rapid increase occurs in later superwind stages when the mass-loss rate reaches $\sim$ 10$^{-4}$ M$_\odot$ yr$^{-1}$. In contrast, super-AGB stars, with their larger radii and higher luminosities\footnote{The surface luminosity of a super-AGB star can be quite extreme, up to $\sim$ 1.5$\times$ 10$^{5}$ L$_\odot$ (M$_{\rm{bol}}$$\sim$ $-8.2$), see Tables~\ref{main} and \ref{main2}.}, reach the superwind phase at, or very near, the start of the TP-(S)AGB phase. They also experience a greater number of thermal pulses, although simultaneously having a reduced TP-(S)AGB phase. This is caused by the substantial shortening of the interpulse period with increasing core mass. 

The \cite{vas93} mass-loss rate prescription is a radiation-pressure-driven limited wind,\footnote{The radiation pressure limit can be approximated as $\dot{M}$=L/c$v_{\rm{exp}}$, were c is the speed of light, L is the stellar luminosity and $v_{\rm{exp}}$ is the stellar expansion velocity.} with the average mass-loss rate during the TP-(S)AGB phase $\langle$$\dot{M}$$\rangle$ increasing with the initial stellar mass and ranging from 0.1$-$1.5$\times10^{-4}$ M$_\odot$ yr$^{-1}$ (Tables~\ref{main} and \ref{main2}). This leads to very short thermally pulsing lifetimes of the order 0.2$-$2$\times10^5$ yr. To put these mass-loss rates in perspective, these values are about a factor of 10 below the observed upper limit for a stable mass-loss rate from a red supergiant or AGB star of $\sim$ 10$^{-3}$ M$_\odot$ yr$^{-1}$ \citep{van99b}. 

The choice of mass-loss rate in super-AGB stars for a range of metallicities has been discussed in considerable detail in previous works in this series. For example, in Paper II (see fig. 7) we analyzed the effect of a range of mass-loss prescriptions (\cite{rei75,blo95,van05,vas93}) for a 8.5 M$_\odot$ Z=0.02 model and found that changes in the final core mass never exceeded 0.03 M$_\odot$ despite the fact that the mass-loss rate varied by over an order of magnitude. This small impact on core mass is a consequence of all standard AGB star mass-loss prescriptions resulting in very rapid mass-loss rates when applied to super-AGB stars, hence any choice of (commonly used) AGB star mass-loss rate will easily dominate the very moderate core growth rate.

\subsection{Low temperature variable composition opacity models}
	
Another key factor which determines the mass-loss rate in super-AGB and massive AGB stars is the envelope opacity.  The use of low temperature molecular opacities which take into account the envelope composition variations during the AGB phase has been shown to be important for cases where the envelope composition ratio C/O exceeds unity. When the star becomes carbon rich, the change in molecular chemistry leads to an increase in opacity which results in a cooler and more extended stellar envelope and a higher mass-loss rate \citep{mar02,cri07,ven10b}. 
In the more metal-rich models (Z=0.004$-$0.02), the stars become carbon-rich only when they have already reached the superwind phase, making the impact of inclusion of updated opacities negligible. This is not the case for the lower metallicity models (Z=0.001 and 0.0001), hence the latter were calculated with the C, N variable composition low-temperature opacities from \cite{led09}. We have not used the most recent treatment of low-temperature opacities from \texttt{AESOPUS} which include C, N \textit{and} O variations \citep{mar09}. \cite{fis14} showed that for models of 5 and 6\,M$_\odot$ at Z=0.001, the variation from C, N to C, N, O variable opacities led to a $\sim$ 30 per cent decrease in TP duration, with the final core masses different by only 0.002$-$0.003 M$_\odot$. This marginal effect is far less significant than, for example, the uncertainties  in the mass-loss prescriptions for super-AGB and massive AGB stars. 

\subsection{Convergence issues at the end of the super-AGB phase - Fe-peak instability}\label{subsec-fe}

\begin{figure}\begin{center}
\resizebox{\hsize}{!}{\includegraphics{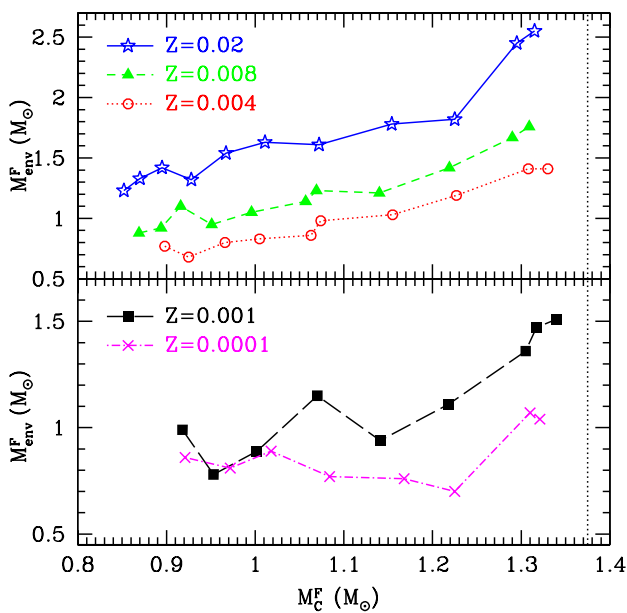}}
\caption{Final remaining envelope mass $M_{\rm{env}}^{\rm{F}}$ at the end of our calculations plotted against the final core mass $M_{\rm{C}}^{\rm{F}}$. This figure is split into two panels based on models with either fixed (top) or variable (bottom) compositional low temperature molecular opacities.} 
\label{mcvsmenv} \end{center}\end{figure}

As commonly mentioned in the literature, the evolution of super-AGB and massive AGB models is terminated due to convergence issues prior to the removal of the entire envelope. 
The remaining envelope when this occurs can be quite massive (up to $\sim$ 2.5\,M$_\odot$), and tends to be more massive as the stellar core (and therefore the initial mass of the model star) increases. The remaining envelope mass when convergence issues ceased our calculations is $M_{\rm{env}}^{\rm{F}}$ and can be found in Tables~\ref{main} and \ref{main2} and in Fig.~\ref{mcvsmenv} as a function of the final core mass $M_{\rm{C}}^{\rm{F}}$ taken at the same time.
From our simulations, we find that lower metallicity models stop converging with lower envelope masses than their more metal rich counterparts. 
The timing of the occurrence of this instability, and its dependence on metallicity, is still debated, with an opposite result found in \cite{wei09}, where metal-rich models evolved to lower envelope masses. Nevertheless, our results are consistent with the explanation of the instability given by \cite{woo86}, which associates the loss of convergence with the condition that just after a late thermal pulse $\beta$, the ratio of the gas pressure to total pressure, tends to zero in shells near the base of the convective envelope, with resultant super-Eddington luminosities. This instability has been attributed to the presence of an Fe-opacity peak near these unstable layers of the star \cite[][]{swe99,lau12}. It is expected that as a consequence of this instability there will be an inflation of the envelope which leads to either its entire ejection or a period of enhanced mass-loss. In either case, further evolution is expected to be substantially truncated. 

The conditions necessary for this Fe-peak instability are intimately related to the temperature gradient near the base of the convective envelope. Thus this phenomenon only occurs in our models when HBB has either ceased or has greatly reduced in efficiency. Observationally this decrease in temperature at the base of the convective envelope near the end of massive AGB star evolution has been inferred from the most luminous Magellanic cloud carbon stars \citep{van99} which \cite{fro98a} suggested to have formed from repeated 3DU events after the HBB has ceased. The temperature at the base of the convective envelope (and therefore part of the conditions required for the occurrence of the instability) is primarily a function of the envelope mass, core mass and metallicity, but is also dependent on the efficiency of mixing (and MLT parameter $\alpha_{\rm{mlt}}$).

Although the Fe-peak is more pronounced at higher metallicities, we suggest that lower metallicity models evolve further, not only due to their lower Fe content, but also for their ability to sustain higher HBB temperatures with less envelope mass due to their more compact structure. 

Recent observational work on OH/IR stars \citep{dev14} have uncovered a potential problem in the current understanding of the mass-loss rates in the later stages of massive AGB star, whereby the superwind phase is considerably shorter than expected.  To remedy this, the authors hypothesised the need for an extremely high and abrupt mass loss phase. We suggest an Fe-peak instability driven mass ejection may offer a potential solution to this problem.
Returning to the effect of the amount of Fe within the envelope, it is of interest to mention the possibility of a critical metallicity below which this Fe-peak instability would not occur. With our lowest metallicity super-AGB star models using the \textsc{monstar} program (Z=10$^{-5}$ \citealt{gil13}), this instability is still encountered. Relating to a similar phenomena, we note that the Fe-opacity peak also drives the $\kappa$ mechanism within $\beta$ Cepheid stars and slowly pulsating B stars \citep{cox92,mos92} with these objects found down to at least the (relatively) low metallicity of the Small Magellanic cloud ($\sim$Z=0.004) \citep{dia08}.

Due to the uncertain nature of the behaviour of the star after the Fe-peak instability, we explore the final fate of these models by performing simple core growth versus mass-loss rate extrapolations to see if the remaining envelope would be removed prior to the model reaching the electron capture core mass limit  $\sim$ 1.375\,M$_\odot$ \citep{nom84,nom87}. We note that this critical value can vary slightly depending on model assumptions, e.g. 1.367\,M$_\odot$ in \cite{tak13}.

For our extrapolations we assume constant mass-loss rate, core growth rate, interpulse period and third dredge-up efficiency, with these inputs obtained by using the average values for the entire (S)AGB phase (taken from Tables~\ref{main} and \ref{main2}) until removal of the entire envelope. We have assumed here that the Fe-peak instability does not increase the subsequent mass-loss rate. Using our method there is at most $\sim$ 0.01\,M$_\odot$ increase in the revised extrapolated core mass $M_{\rm{C}}^{Ex}$ for all models which ceased computation due to Fe-peak instability. We note that our extrapolations will lead to an over estimate of the final core mass in the lower mass models, where the average value of $\dot{M}$ is significantly lower by about a factor of 10 than the mass-loss rate near the end of the calculations. Whilst we could have performed more sophisticated extrapolations, we feel this was not warranted because no significant increase in accuracy would be achieved due to the large uncertainty to the nature or existence of the Fe-peak opacity instability.
Furthermore, as the core masses within these above-mentioned models are not very close to the limit for electron-captures, this small variation in core mass due to the extrapolations does not change their final fate: they will remain as either CO, CO(Ne) or ONe WDs.  

Unfortunately another numerical issue plagued computations and stopped further evolution in the most massive models undergoing thermal pulses when core masses exceeded $\sim$ 1.33$-$1.35\,M$_\odot$. The cause of this issue is currently under investigation, but is not related to either the Fe-peak instability, or to the onset of electron-capture reactions within the core as the density is still far below the critical values. For the models in which this occurred the final envelope masses are denoted in italics in Tables~\ref{main} and \ref{main2}. Here we also extrapolated these models to determine their possible outcomes. During these extrapolations the core growth is larger $\sim$ 0.02 \,M$_\odot$, but just as with the Fe-peak instability models, the final extrapolated core masses are below the electron-capture limit, hence these stars also would end life as ONe WDs. We draw attention to the 9.3\,M$_\odot$ and 9.4\,M$_\odot$ Z=0.008 models because their extrapolated core mass values are very close to the electron-capture limit, so with a slightly different extrapolation scheme they could reach conditions for an explosion. However, most likely these models would have reached the conditions for the Fe-peak instability so they would also end their lives as ONe WD.   

\section{Final Fates}\label{sec-ff}

\begin{figure*}
\begin{center}
\includegraphics[width=11cm,angle=0]{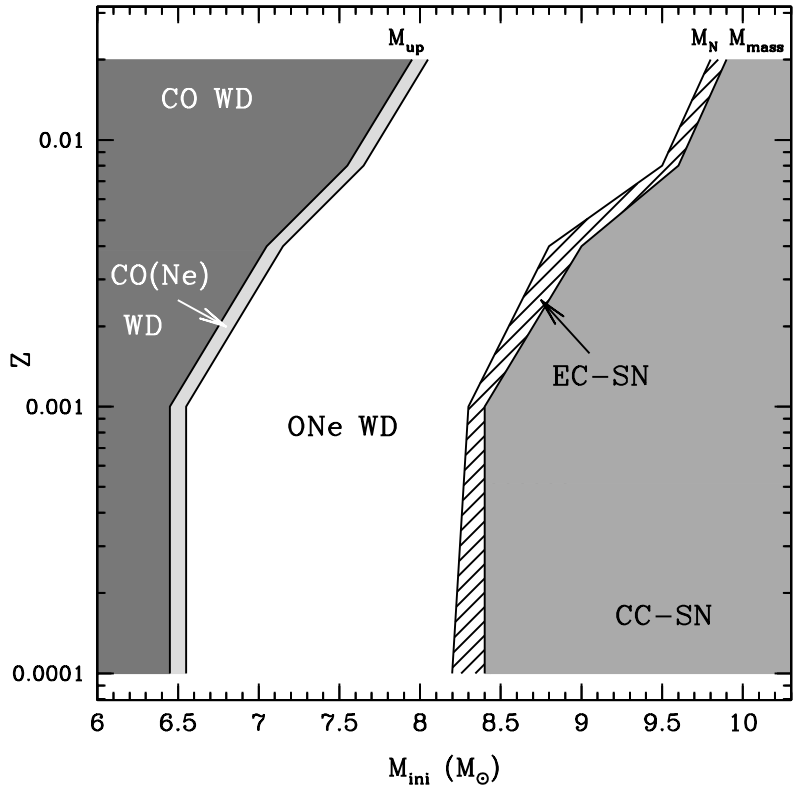}
\caption{Final fates of intermediate-mass stars. Solid lines delineate $M_{\rm{up}}$, $M_{\rm{n}}$ and $M_{\rm{mass}}$. The hatched region represents our suggested maximum width of the EC-SN channel. We also draw particular attention to the hybrid CO(Ne) WD region.}
\label{shade}\end{center}
\end{figure*}

Fig.~\ref{shade} gives a global overview of our results in mass and metallicity space. The critical boundary values of $M_{\rm{up}}$, $M_{\rm{n}}$ and $M_{\rm{mass}}$ are also shown in this Figure and reported in Table~\ref{tab-crit}. 

Our models with moderate mass-loss rates and efficient 3DU increase their core mass by only $\approx$ 0.01$-$0.03\,M$_\odot$ during the TP-(S)AGB phase. This can be seen by comparing the post- 2DU core mass $M_{\rm{2DU}}$ to the final core mass M$_{\rm{C}}^{F}$ in Tables~\ref{main} and \ref{main2}. 

\subsection{White Dwarfs - CO, CO(Ne), ONe}\label{subsec-wd}

\begin{figure}\begin{center}
\resizebox{\hsize}{!}{\includegraphics{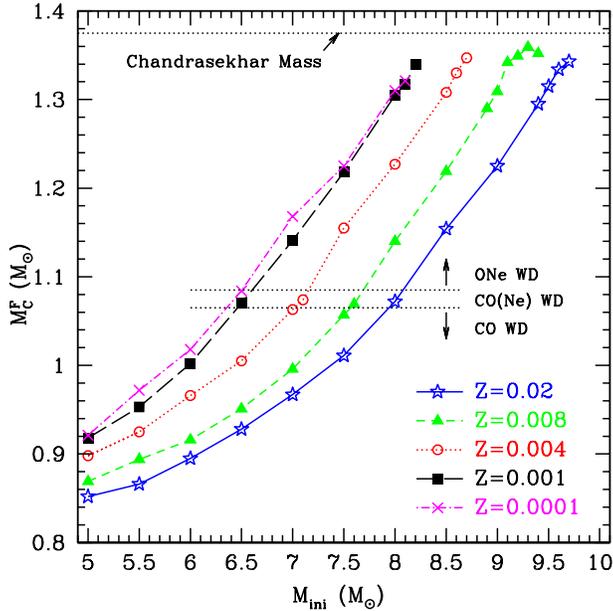}}
\caption{Initial to Final Mass relation, when convergence issues cease calculations. The dotted horizontal lines that delineate the difference types of WD are illustrative only.}
\label{IFMR}\end{center} \end{figure}

In Fig.~\ref{shade} the mass boundaries of the three types of massive white dwarfs, CO, CO(Ne) and ONe, are shown. We draw particular attention to the hybrid CO(Ne) WDs, which in our calculations occupy a thin initial mass range (of width $\sim$ 0.1$-$0.2\,M$_\odot$) with final core masses in the range 1.06$-$1.08\,M$_\odot$ (Fig.~\ref{IFMR}). If one assumes a \cite{kro93} IMF one would still expect them to be quite numerous, making up about 6$-$8 percent of WDs that have undergone either complete or partial carbon burning. Unfortunately direct observations of these hybrid CO(Ne) WDs would be impossible due to their outer shell of CO which would make them indistinguishable from a massive CO WD. Nevertheless, confirmation of their existence may come indirectly via the characteristic of the light curves of Type Ia SN explosions, if they belong to a close binary system \citep{den13a}.  We would also expect differences in their cooling behaviour, because WD cooling is strongly dependent on composition \cite[e.g.][]{sal97,sal10,alt07}. 

The minimum mass of an ONe WD is $\sim$1.06\,M$_\odot$, in close agreement with the estimates found in previous studies \cite[e.g.][]{pap1,sie06}. The value of $M_{\rm{up}}$ range between 6.5 and 8.0\,M$_\odot$ (Table~\ref{tab-crit}), and decrease with decreasing metallicity. Compared to our earlier work (Paper I) we find a downwards shift in $M_{\rm{up}}$ of $\approx$ 1\,M$_\odot$ for each metallicity, as a consequence of our different treatment of the convective boundaries (see Sect.~\ref{sec-Super AGB}).

The initial mass range for production of ONe WDs has a relatively constant (albeit at differing values) width of about 1.5\,M$_\odot$ over the range of metallicities analyzed in this work. A substantial numbers of massive and ultra-massive WDs\footnote{Massive WDs are deemed to be those greater than 0.8\,M$_\odot$ whilst ultra-massive WDs have masses in excess of 1.1\,M$_\odot$.} have been discovered in white dwarf surveys; with estimates that about 20 per cent of all WDs are massive WDs \citep{lie05,kep07} and about 10 per cent ultra-massive WDs \citep{ven08}. We note that a second WD mass distribution peak at $\sim$ 1.04\,M$_\odot$ was found by \cite {nal04} which interestingly closely corresponds to the divide between CO and ONe core compositions (Fig.~\ref{IFMR}). 

Possible observational evidence for ONe WDs come from the study of \cite{gan10}, where two O-rich WDs were discovered with masses $\approx$ 1\,M$_\odot$. ONe WDs are also suspected to host neon nova \citep{jos98,wan99} and may be quite common \citep{gil03}.
The surface composition of neon nova can be used as a thermometer to indirectly determine the underlying ONe WD masses \citep{wan99}. Recently \cite{dow13} used this approach to determine the WD mass for V838 Her to be $\sim$ 1.34$-$1.35\,M$_\odot$. 
\cite{weg12} used data of Galactic WDs to show that the kinematics of the majority of high-mass WDs with $M_{\rm{C}}$ $>$ 0.95\,M$_\odot$ are consistent with being formed through a single star evolution channel. 

In Fig.~\ref{IFMR} we provide a theoretical initial to final mass relation (IFMR) for the time when convergence issues have ceased calculations. This figure shows that stars of lower metallicity produce more massive cores for the same initial mass. For the two lowest metallicity models the IFMR is very similar which is related to the comparable core mass during the core helium burning phase (refer Section~\ref{sub-2du}).
One can notice that the slope of these curves is similar to those depicted in Fig.~\ref{fig-post2du} showing the core mass after the 2DU versus initial mass. The preservation of this slope results from the fact that both massive AGB and super-AGB stars have similar effective core growth rates. The higher 3DU efficiency of massive AGB stars is compensated by the fact that they have slower wind mass loss rates and hence longer TP-AGB evolution than their more massive super-AGB stars counterparts.

\begin{figure}
\begin{center}
\resizebox{\hsize}{!}{\includegraphics{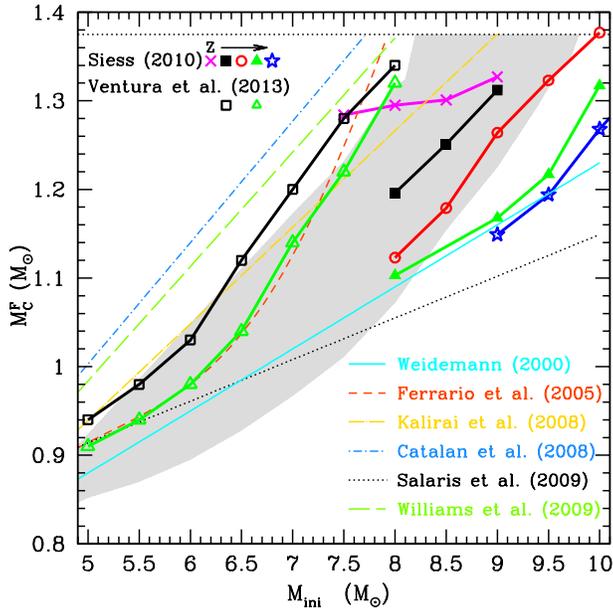}}\end{center}
\caption{Initial to Final Mass relation. The shaded region represents the boundary of the values for our calculations (from Fig~\ref{IFMR}). We have included a large selection of empirical and semi-empirical IFMRs as indicated in the legend. The symbols and thick lines represent the theoretical results from the calculations by \protect\cite{sie10} and \protect\cite{ven13}.}
\label{fits} \end{figure}

In Fig.~\ref{fits} we superimpose a selection of the most commonly used empirical/semi-empirical IFMRs over our theoretical IFMRs. The IFMRs compared are those by \cite{wei00}, \cite{fer05}, \cite{cat08}, \cite{wil09}, \cite{kal08} and \cite{sal09}, with the majority of these observationally based IFMRS derived from studies of WDs within star clusters. Clearly noticeable in this figure is the large disparity in the maximum initial mass which forms a WD with values ranging from $\sim$ 7.7\,M$_\odot$ to greater than 10.0\,M$_\odot$. The large variation between prescriptions is due to a variety of causes, including uncertainties in determinations of cluster age, metallicity, as well as WD cooling ages. Our models do not seem to favour the traditional linear form of the IFMR but correspond more closely, at least qualitatively, to the shape of the IFMR from \cite{fer05}. 

In Fig.~\ref{fits} we also compare our results to the IFMR obtained from detailed super-AGB star calculations by \citeauthor{sie10} (2010, hereafter S10) for metallicities Z=0.0001$-$0.02 and super- and massive AGB star calculations by \citeauthor{ven13} (2013, hereafter V13) for metallicities Z=0.001 and 0.008. Significant differences in the IFMRs are evident between these previous studies and this current work. The differences can be divided into two main characteristics, the (initial) mass range and the slope.

The initial mass range of the IFMR for massive/ultra-massive WDs is primarily related to the determination of convective boundaries during the core helium burning phase. Stellar models which utilized the strict Schwarzschild criterion \citep{pap5,sie07} have smaller core masses than overshoot models and consequently will have their IFMR shifted towards higher initial masses.
Because the Schwarzschild approach was used in the S10, his model have smaller final core masses for the same initial masses when compared to our study. The V13 models on the other hand have larger final core masses for the same initial masses due to their implementation of convective overshooting during both core H and He burning phases\footnote{These models assume an exponential decay of convective velocities starting from convective boundaries, with the e-folding distance given by $\zeta$$H_P$ with $\zeta$ =0.02 \citep{ven13}.} which produces larger core masses than our models based on the search for convective neutrality.

The slope of the IFMR is influenced by multiple factors, in particular the efficiency of 2DU/dredge-out and the effective core growth rate as discussed in detail in Section~\ref{sec-tp}. 
Even though there is no 3DU in the V13 calculations, the rapid mass-loss (\citealt{blo95} with $\eta$=0.02) only allows for a modest increase in the core mass along the TP-(S)AGB phase of $\sim$ $0.02-0.03$\,M$_\odot$. Another feature of their calculations is a more efficient core growth rate, with a rough estimate suggesting rates $\approx$ 2$-$5 times faster than our calculations, most likely caused by their treatment of convection based on the full spectrum turbulence approach \citep{can91}. Within this formalism the temperature at the base of the convective envelope is higher, making the H burning shell advance faster than in the standard MLT case. 

In the S10 calculations 3DU is also absent, however as those models were computed with the moderate \cite{vas93} mass-loss prescription, the increase in the core mass is larger than in V13, of the order of 0.06$-$0.08\,M$_\odot$.  
Because there is no surface metallicity enrichment from 3DU events in the S10 model calculations, the stars retain a more compact structure, which favours lower mass-loss rates and hence longer TP-(S)AGB phases and more core growth resulting in a flatter IFMR relation. This is particularly noticeable at lower metallicities where the pollution of the envelope by 3DU events can significantly impact the mass loss rate.

Even though the input physics varies considerably between these three detailed evolutionary studies, which has resulted in a large disparity in the number of thermal pulses, duration and core growth rates in the TP-(S)AGB phase, their is an overall consensus between these results. Namely that the vast majority of stars that enter the thermally pulsing super-AGB phase will end life as ONe WDs. 

\subsection{$M_{\rm{n}}$ - Neutron star boundary}

All of our models that ignite carbon and experience a 2DU or dredge-out episode are able to enter the TP-(S)AGB phase and expel their envelopes before their core masses reach $M_{\rm{CH}}$.  
Therefore, the only mass range which could result in EC-SN from our study are those models (denoted in bold in Tables~\ref{main} and \ref{main2}) that reach Ne ignition condition with core masses which were between 1.35\,M$_\odot$$\la$ $M_{\rm{C}}$ $\la$ 1.37\,M$_\odot$ near the end of carbon burning.  Our calculations indicate that the EC-SN channel is limited to a very narrow initial mass range of $\sim$ 0.1$-$0.2\,M$_\odot$ width delineated in Fig.~\ref{shade} by the hatched region  between $M_{\rm{n}}$ and $M_{\rm{mass}}$.

Using a \cite{kro93} initial mass function (IMF), we calculate the percentage of EC-SNe compared to the total number of gravitational collapse supernovae. Our results, provided in Table~\ref{tab-crit} show that the EC-SN channel makes up less than 5 per cent of all gravitational collapse supernovae in the metallicity range studied here. 

Because we do not explicit apply a metallicity scaling to our mass-loss rates we do not find the rate of EC-SN to be significantly higher at lower metallicities. Another consequence of this assumption is that CO cores are unable to grow massive enough during the TP-AGB phase to explode as Type 1.5 SN. This is in agreement with our lower metallicity Z=10$^{-5}$ study by \cite{gil13}.

\begin{table}
\caption{Critical Boundary Masses: $M_{\rm{up}}$, $M_{\rm{n}}$ and $M_{\rm{mass}}$. Assuming a \protect\cite{kro93} IMF, the percentage of EC-SNe as a fraction of the total gravitational collapse supernova rate from our study and from the favoured parametrised values from \protect\cite{poe07}.} \label{tab-crit}
 \begin{center} 
\setlength{\tabcolsep}{2.5pt}
\begin{tabular}{lcccccccc}
\hline
Z&$M_{\rm{up}}$&$M_{\rm{n}}$&$M_{\rm{mass}}$&\multicolumn{2}{c}{EC-SN $\%$} \\
 &(M$_\odot$)&(M$_\odot$)&(M$_\odot$)&This study &P07 \\
\hline
0.02     &8.0& 9.8 &9.9 & 3.5 & 3     \\ 
0.008    &7.6& 9.5 &9.6 & 2   & 9     \\
0.004    &7.1& 8.8 &9.0 & 3   & 11    \\
0.001    &6.5& 8.3 &8.4 & 3.5 & 12.5  \\ 
0.0001   &6.5& 8.2 &8.4 & 5   & 21    \\ 
\hline
\end{tabular}
\end{center}
\end{table}

We compare our results to the very detailed EC-SN parametric studies by \citeauthor{poe07} (2007, hereafter P07\footnote{Their solar metallicity (Z=0.02) case was presented in \protect\cite{poe08} whilst an overview for their entire metallicity range provided in fig.12 in \protect\cite{lan12}}) and \citeauthor{sie07} (2007, hereafter S07). These works span the metallicities considered here. By the virtue of synthetic studies they produced a series of parametrisations but here we compare to their `best estimate' or favoured result. 

In S07 the core masses at the start of the thermally pulsing phase were calculated from detailed stellar models then an extrapolation was performed using a parameter $\zeta$, the ratio of mass lost from the envelope to the effective core growth rate. Larger values of $\zeta$ result in a finer EC-SN channel. With the adopted values of $\zeta$ between $\sim$ 35-100, a relatively constant EC-SN window of $\approx$ 0.8$-$0.9\,M$_\odot$ was found, resulting in $\sim$ 15 per cent of all SN being EC-SN. \footnote{There is a a typographically error in \protect\cite{sie07} pg.905, where the $\dot{M}_{\rm{env}}$ should be 5$\times$10$^{-5}$\,M$_\odot$ yr$^{-1}$ instead of 5$\times$10$^{-4}$\,M$_\odot$ yr$^{-1}$.} If we use however, values representative from our models, we attain a value of $\zeta$ $\sim$ 300, which corresponds to the fine EC-SN window of 0.2\,M$_\odot$ in initial mass.

In P07 post 2DU core masses were obtained by interpolating in a grid of massive and intermediate-mass models. Then a synthetic extrapolation routine was used to determine the final fate. Their standard model assumed the mass-loss rate from \cite{van05} with a metallicity scaling $\sqrt{Z/Z_\odot}$ from \cite{kud87}, used the core growth rate taken from detailed models and parameterised efficient 3DU. With this combination of input physics the width of the EC-SN channel ranged from 0.2$-$1.3\,M$_\odot$. This corresponds to EC-SNe comprising 3$-$21 per cent of all SN events, with this value larger at lower metallicity (Table~\ref{tab-crit}). With similar core growth rates, 3DU efficiencies, as well as the standard mass-loss rates in our study and P07, the main difference between our results is the effect of the metallicity scaling upon the mass-loss rate.  As discussed in section~\ref{subsec-ml}, the application of a metallicity scaling upon the mass-loss rate of AGB stars may be limited because of envelope enrichment by either corrosive 2DU, dredge-out, or 3DU (if present). 

If we assume a hypothetical situation where all ONe WDs in our study grew massive enough to explode as EC-SNe, then $\approx$ 25$-$27 per cent of all SNe would come from this channel, putting an upper limit on the number of supernova from super-AGB and hyper-AGB stars. This value compares well with the $\approx$ 20 per cent calculated by \cite{poe08} for their study at solar metallicity.

\section{Conclusions}

We have explored the fate of super-AGB and massive AGB stars using detailed stellar evolutionary calculations including the thermally pulsing phase. 
We provide an initial to final mass relation for massive and ultra-massive WDs as well as the mass range for the hybrid class of WD, the CO(Ne)s.

Our models are characterized by a high efficiency of 3DU which combined with moderate mass loss leads to only a modest core growth of $\sim$ 0.01$-$0.03\,M$_\odot$ during the thermally pulsing (super)AGB phase. Due to this, the majority of our computed models will end their lives as either CO, CO(Ne) or ONe WDs. In our study only stars that are massive enough to reach neon burning conditions near the end of carbon burning will grow enough to reach the Chandrasekhar mass and undergo an EC-SN. 
From this we conclude the EC-SN channel (for single stars) is very narrow in initial mass, at most $\approx$ 0.2\,M$_\odot$. This corresponds to between $\sim$ 2$-$5 per cent of all gravitational collapse supernovae being EC-SNe in the metallicity range Z=0.02 to 0.0001. 

Our results also show that the EC-SN channel is \textit{not} significantly wider in initial mass for lower metallicity stars. This is based on the assumption that mass-loss rates are not fundamentally different for (initially) low metallicity stars, once they have been substantially enriched in ``metals'' during their lifetimes from dredge-up or dredge-out events. The most massive super-AGB stars will become carbon-rich either from dredge-out or corrosive 2DU events which leads to higher envelope opacity due to changes in molecular chemistry and enhanced mass loss. The recent work by \cite{con14} found that this effect is important even for primordial AGB stars. This could lead to an interesting situation at the lowest (primordial) metallicities, where models that are enriched prior to the TP-SAGB phase from dredge-out events or corrosive 2DUs could lose their envelope mass while on the AGB and end their lives as ONe dwarfs whilst initially less massive models which do not undergo enrichment could explode as either EC-SNe or a Type 1.5 SNe due to slower mass loss. Even if this effect is not as dramatic as to change the final fate of primordial AGB star models, it may well alter their amount of core growth and ultimately cause a possible non-linearity of the initial to final mass relation.

Although the mass-loss rate, efficiency of 3DU and convective boundary approach during core helium burning are clearly important in determining $M_{\rm{n}}$, we suggest the 2DU/dredge-out behaviour and interplay between the He, C and Ne burning shells, and mixing process near the end of the carbon burning phase are arguable the most important factors (and also unfortunately the most code modelling dependent) in determining the occurrence and or width of the EC-SN channel.

We suggest dredge-out events, with their potential heavy element production, may provide a unique (and observable) signature to distinguish the differing evolutionary histories near the end of the carbon burning.

The Fe-peak instability which halts calculations prior to expulsion of the entire envelope may impact (reduce) the width of the EC-SN channel but by how much will remain an open question until hydrodynamical simulations of this type of event become available. 

\section{Acknowledgements}
This research was supported under Australian Research Council’s Discovery Projects funding scheme (project numbers DP0877317, DP1095368 and DP120101815) and the Go8-DAAD Australia/Germany Joint research cooperation scheme. LS is a FNRS Research Associate. CLD would like to thank  A.J.T. Poelarends, J.J. Eldridge and R.J. Stancliffe for interesting discussions.\\

\bibliographystyle{mn2e} 
\bibliography{FFMnras} 

\end{document}